\newcommand{\Pl}{\partial}
\newcommand{\ts}{\textstyle}
\newcommand{\bee}{\begin{equation}}
\newcommand{\ene}{\end{equation}}
\newcommand{\beea}{\begin{eqnarray}}
\newcommand{\enea}{\end{eqnarray}}

\newcommand{\fdar}[2]{\frac{{\ts d \/ #1}}{{\ts d\/ #2}}}

\newcommand{\fpar}[2]{\frac{{\ts \Pl \/ #1}}{{\ts \Pl \/ #2}}}
\newcommand{\nder}[3]{\frac{{\ts d^{#1} \/ #2}}{{\ts d \/ #3^{#1}}}}

\documentclass[aps,prb,preprint]{revtex4-1}
 \usepackage{graphicx}
 \usepackage{dcolumn}
 \usepackage{bm}
 \usepackage{latexsym}
 \usepackage{amsmath}
 \usepackage{amssymb}
 \usepackage{amsfonts}
 \usepackage{amsthm}
 \begin{document}
 \title{Kelvin-Helmholtz Instability in non-Newtonian Complex Plasma}
 \author{D. Banerjee, S. Garai, M. S. Janaki and N. Chakrabarti}
 \affiliation{ Saha Institute of Nuclear Physics,
  1/AF Bidhannagar, Kolkata - 700 064, India}
\begin{abstract}
     The Kelvin-Helmholtz (KH) instability is studied in a non-Newtonian dusty plasma with an experimentally verified model [Phys. Rev. Lett. {\bf 98}, 145003 (2007)] of shear flow rate dependent viscosity. The shear flow profile used here is a  parabolic type bounded flow. Both the shear thinning and shear thickening properties are investigated in compressible as well as  incompressible limits using a linear stability analysis. Like the stabilizing effect of compressibility on the KH instability, the non-Newtonian effect in shear thickening regime could also suppress the instability but on the contrary, shear thinning property enhances it. A detailed study is reported on the role of non-Newtonian effect on KH instability with conventional dust fluid equations using standard eigenvalue analysis.
\end{abstract}
\maketitle

\section{Introduction}
\label{intro}
Complex plasma is characterized by the presence of the micron sized charged dust particles in a normal electron-ion plasma. These type of systems are greatly affectionate to the plasma physics community due to their natural occurrence in different places in our universe i.e. in planetary rings, cometary tails, white dwarf matter and interstellar clouds etc. It also has existence in human made systems like  plasma processing and plasma etching equipments in industry, fusion devices, rocket exhausts etc and for such wide occurrence it is important to characterize the different properties and novel features of the dusty plasma.
Since the macroscopic dust particles can be visualized and tracked at particle level, dusty plasma has been treated as a good experimental medium  to study phase transition\cite{morf}, transport properties\cite{ratn,liu}, crystal formation\cite{manish,htho} and other collective phenomena\cite{piep,kmp}. The neutral dust particles when inserted  into a laboratory plasma, become highly charged due to the different charging mechanism like plasma currents, photoelectric effects, secondary emission etc.  Due to the higher mobility of the electrons than the ions, the dust particles acquire high negative charge. In dusty plasma experiments, charged micro particles levitate in the sheath at the lower electrode in gas discharge and it forms Yukawa systems where the interaction potential between the particles is of the form $\phi(r)\propto Q $exp$(-r/\lambda_D)/r$, ($r$ is the separation between two particles and $Q$ is the charge of each particle) which express Coulomb repulsion that is exponentially suppressed with screening length $\lambda_D$. When temperature exceeds melting temperature Yukawa system shows liquid state of matter and generation of shear flow in such configuration has enabled the measurement of the shear viscosity. In such system Nosenko and Goree\cite{nose} measure shear viscosity by using two parallel but counter propagating laser beams to generate a shear flow in a planar Couette configuration. Using molecular dynamics simulation in a  2D Yukawa liquid, Liu and Goree\cite{binl} have reported the dependence of shear viscosity on the temperature of random thermal motion of dust particles through Coulomb coupling parameter $\Gamma$. In this context, we should mention of the simulation works which have predicted the signature of non-Newtonian property\cite{dnk}.
Investigations of shear flows in a complex plasma fluid by Ivlev et. al\cite{ivle} have revealed  the signature of non-Newtonian property with the viscosity coefficient varying with the velocity gradient. The experiment has been done  with  gas induced shear flow for different discharge currents and also by applying laser beams of different power. This has enabled measurement of the shear viscosity and confirmation of the non-Newtonian property over a considerable range of shear rates. Gavrikov et al. have also reported\cite{jpps} this phenomenon in a dusty plasma liquid.

In fluid systems, it is well known that the inhomogeneous bounded shear flow  can drive  the KH instability which is widely studied in experiment and also theoretically investigated with the famous Orr-Sommerfeld equation in twentieth century \cite{kundu,rpnt}. In an inviscid parallel flow without any point of inflection in the velocity profile (like parabolic flow), the disturbance field cannot extract energy from the basic shear flow,  resulting in stable flow, but onset of viscosity makes it eligible for drawing energy and hence viscosity could destabilize such flow. The non-Newtonian property in complex plasma shows both shear thinning and thickening property depending upon the values of shear rate and parameter regime, so it is expected that these properties may have the opposite effect on the KH instability of inhomogeneous parabolic type profile. Motivated by these ideas, we have studied the effect of the velocity shear rate dependent viscosity on the growth rates and its dispersion by using the standard matrix eigenvalue technique.

This paper is organized in the following manner: Section(II) states the system and its equilibrium. The equilibrium momentum equation is solved numerically to get the equilibrium velocity profile and corresponding non-Newtonian viscosity which is also a function of space through dependance with velocity shear rate. Section(III) contains the derivation of the associated linear equations.   Section(IV) contains the description of the nonlocal analysis including numerical procedure for obtaining the eigenvalues.  The results showing the effect of the shear thinning and thickening property on the KH instability have been presented in this section.  Finally a conclusion is drawn in section(V).

\section{System and its Equilibrium} \label{sec:bas}
In discharge plasma, the dust particles forming dust cloud levitate vertically (z-direction) in presence of external vertical electric field which balances the gravity of the dust particles. A bounded equilibrium flow is generated along the axis of the cylindrical vessel (y-direction)  with variation in the perpendicular x-direction. In our analysis, we consider the flow region as a slab ($-L < x < L$) with the maximum flow speed in the middle of the discharge tube ($x = 0$) and velocity vanishes along the boundary. In such an inhomogeneous charged dust flow, a small wavy disturbance could be unstable  which leads to the well known KH instability. Due to the non-Newtonian property of the dusty plasma, the unperturbed flow would deviate from the parabolic shape. Non-Newtonian viscosity has specific functional dependence on equilibrium shear flow rate and for analytical purpose proper mathematical model is required in this context. In case of complex plasma, the experimentally verified model for the kinematic viscosity $\nu(\gamma)$ with shear rate $\gamma$, given in  Ref.\cite{ivle} can be written as
 \bee
\nu(\gamma)=\frac{2(1+\epsilon)}{\sqrt{1+4\gamma^2-4\epsilon \gamma^4}+1-2\epsilon\gamma^2}\bar{\nu},
\label{model}
\ene
where $\bar{\nu}$ is the value of newtonian viscosity, $\gamma$ is  equilibrium velocity shear rate defined as $\gamma=dv_0/dx$ which is normalized by $({\beta v^2_{T_0}/\bar{\nu}})^{1/2}$. Here $\beta $ is the friction rate, $v^2_{T_0}$ is the thermal velocity. The
other parameter $\epsilon$ which characterizes the non-Newtonian property is given as
 $\epsilon = ({\cal A}/{\cal B}) \left(T_0/T_m\right)^{\alpha+\tau}$ and $\alpha=\tau=1$ as indicated in \cite{ivle}. Here, $T_0$ is the temperature at zero shear rate and $T_m$ is the melting temperature and $\cal A, \cal B$ are weal function of density
 as given in the above mentioned reference. Since we are interested to study fluid properties of complex plasma, in our analysis  $T_0>T_m$ \cite{saigo}. In the limit $\gamma,\epsilon \rightarrow 0$, the model converges to the Newtonian viscosity limit $\nu \rightarrow \bar{\nu}$.

Weakly coupled unmagnetized dusty plasma is completely described by the three basic equations (the continuity equation obeying the mass conservation, the Navier-Stoke's equation showing the momentum balance and the Poisson's equation which connects the potential fluctuation with the density variation) which are the following:
  \bee
  \fpar{n}{t} + \nabla \cdot (n {\bf v}) = 0,
  \label{continuity}
  \ene
  \bee
  \rho \left(\fpar{}{t} + {\bf v}\cdot\ \nabla \right) {\bf v} +n e Z  {\bf E} + c_{d}^{2} \nabla \rho = \fpar{\sigma_{ij}}{x_j},
  \label{momentum}
  \ene
  \bee
   \nabla \cdot {\bf E} = 4 \pi e(n_e + Z n -n_i),
  \label{poission}
  \ene
where $c_{d}=\sqrt{T_d \mu_d \gamma_d/m}$, $T_d$ is the dust temperature due to random thermal motion, $\mu_d$ and $\gamma_d$ are respectively compressibility factor and adiabatic index \cite{pkaw}, $m$ is dust mass.The electric field is denoted as ${\bf E}$, $n$ is defined as dust number density where mass density $\rho = n m$, ${\bf v}$ is the dust fluid velocity and $Z$ denotes number of electronic charge on dust particle.
 Here, we consider low neutral gas pressure so that dust neutral collision becomes less effective and may be neglected. Usually, by colliding with neutrals, dust particles lose their free energy essential for the instability and hence it leads to collisional damping.
We are interested to study low frequency wave ($\omega \ll kv_{Te}, kv_{Ti}$, where
 $v_{Te}, V_{Ti}$  are thermal velocities of electrons and ions respectively) hence the electron and ion dynamics are considered to obey the Boltzmann relation.  For electrostatic mode ($\bf E=-\nabla \varphi$), the electron density ($n_e$) and the ion density($n_i$) are connected with the electrostatic potential($\varphi$) as,
\bee
n_e = n_{e0} \exp(e\varphi/T_e), ~~~~~~~~~~~~~~~~~ n_i = n_{i0} \exp(-e\varphi/T_i),
\label{poiss}
\ene
where $T_e$ and $T_i$ represents electron and ion thermal temperature measured in the Boltzmann unit, $n_{e0}$ and $n_{i0}$ are density of electron and ion fluid at zero potential.
The viscous stress tensor is expressed as\cite{stei},
\[
  \sigma_{ij} = \eta(\gamma)\left[ \left( \fpar{v_i}{x_j} + \fpar{v_j}{x_i}\right) - \frac{2}{3} \delta_{ij} \left( \nabla \cdot {\bf v}\right) \right],
\]
where $\eta(\gamma)$ is an non-Newtonian viscosity coefficient and effect of bulk viscosity is not considered. In ($x-y$) plane $\sigma$ is a $2 \times 2$ matrix with elements $\sigma_{xx}$, $\sigma_{xy}$, $\sigma_{yx}$ and $\sigma_{yy}$.
The coefficient of shear viscosity $\eta$ is  constant for a Newtonian fluid but for non-Newtonian fluid it takes the form $\eta(\gamma)$ where $ \gamma = \sqrt{ II/2}$. The second scalar invariant of rate-of-strain tensor\cite{emit} is of the form
\bee
II = \sum_i \sum_j \left(\frac{\partial v_{i}}{\partial x_{j}} + \frac{\partial v_{j}}{\partial x_{i}}\right)\left(\frac{\partial v_{i}}{\partial x_{j}} + \frac{\partial v_{j}}{\partial x_{i}}\right)
\nonumber
\ene
where suffices $i,j$ varies as $x,y$.


In equilibrium, the density and temperature are assumed to be constant and a constant electric field ($E_0$) is directed along the y-direction. Dust particles are drifted along the y-direction with bounded equilibrium velocity profile having inhomogeneity along the x-direction (perpendicular to the electric field). For homogeneous density and pure shear flow, the continuity equation is automatically satisfied,  and the Poisson's equation leads to quasi-neutrality of the system i.e. $(n_{e0} + Z n_0 -n_{i0}=0)$. From eq.(\ref{momentum}), the y-component of the equilibrium momentum equation can be written as
 \bee
 \fdar{}{x}\left[ \eta_0(\gamma) \gamma\right] = e Z n_0 E_0,
 \label{eqlb}
 \ene
 where $\gamma = dv_0/dx$. In the above equation subscript `$0$' indicates equilibrium quantities. In experiment\cite{ivle}, gas-induced flow is used to generate equilibrium shear flow of dust particles. To take into account this experimental condition following  ref\cite{ivle} we have included
  another space dependent term $A x^2 $ in RHS of Eq. (\ref{eqlb})  where $A$ is the gas drag coefficient.
  Therefore Right hand side of Eq.(\ref{eqlb}) now would be  $ F_0(x)=A x^2 + e Z n_0 E_0$.
\begin{figure}
  \begin{center}
    \begin{tabular}{cc}
      \resizebox{80mm}{!}{\includegraphics{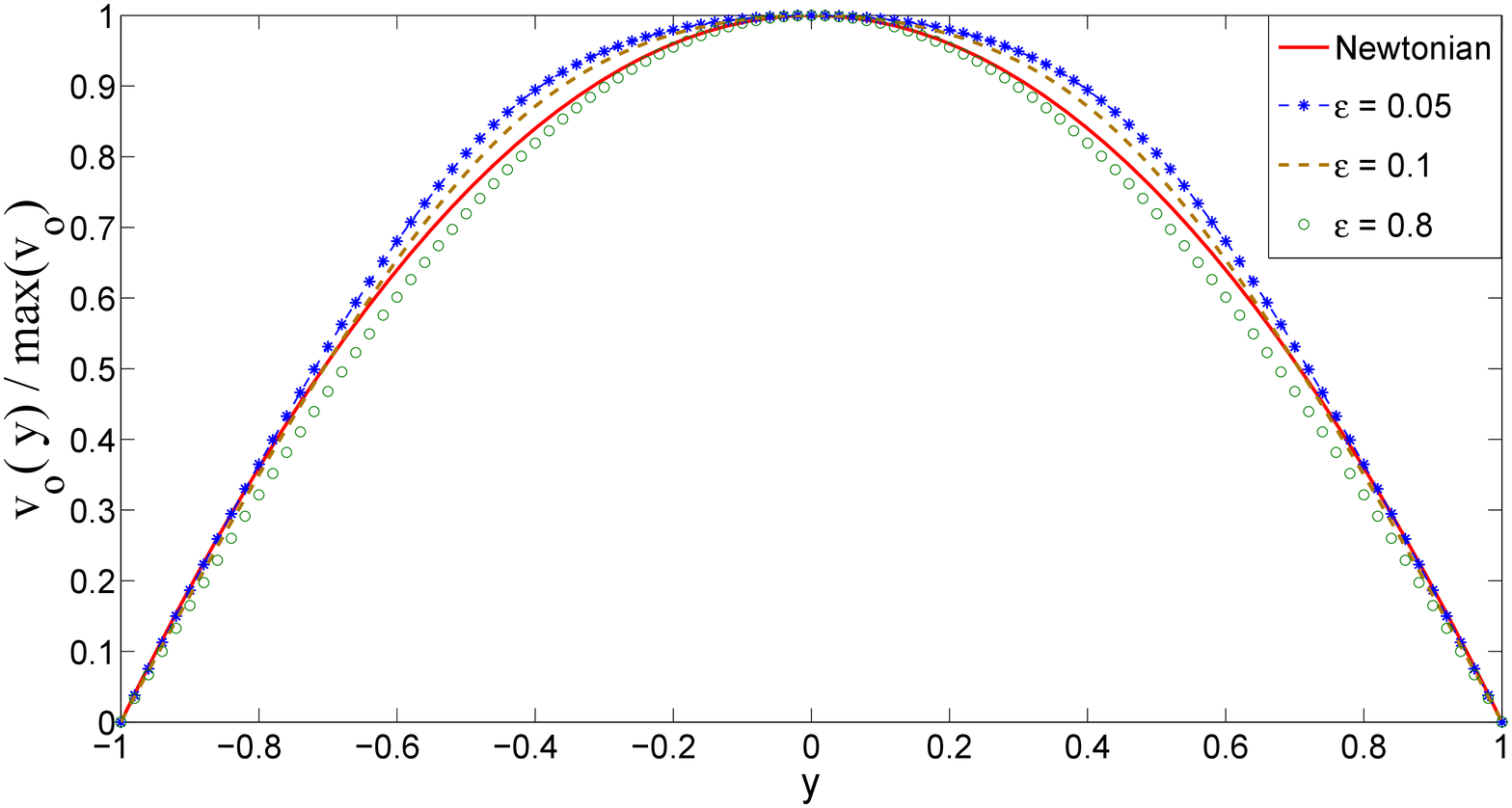}} &
      \resizebox{85mm}{!}{\includegraphics{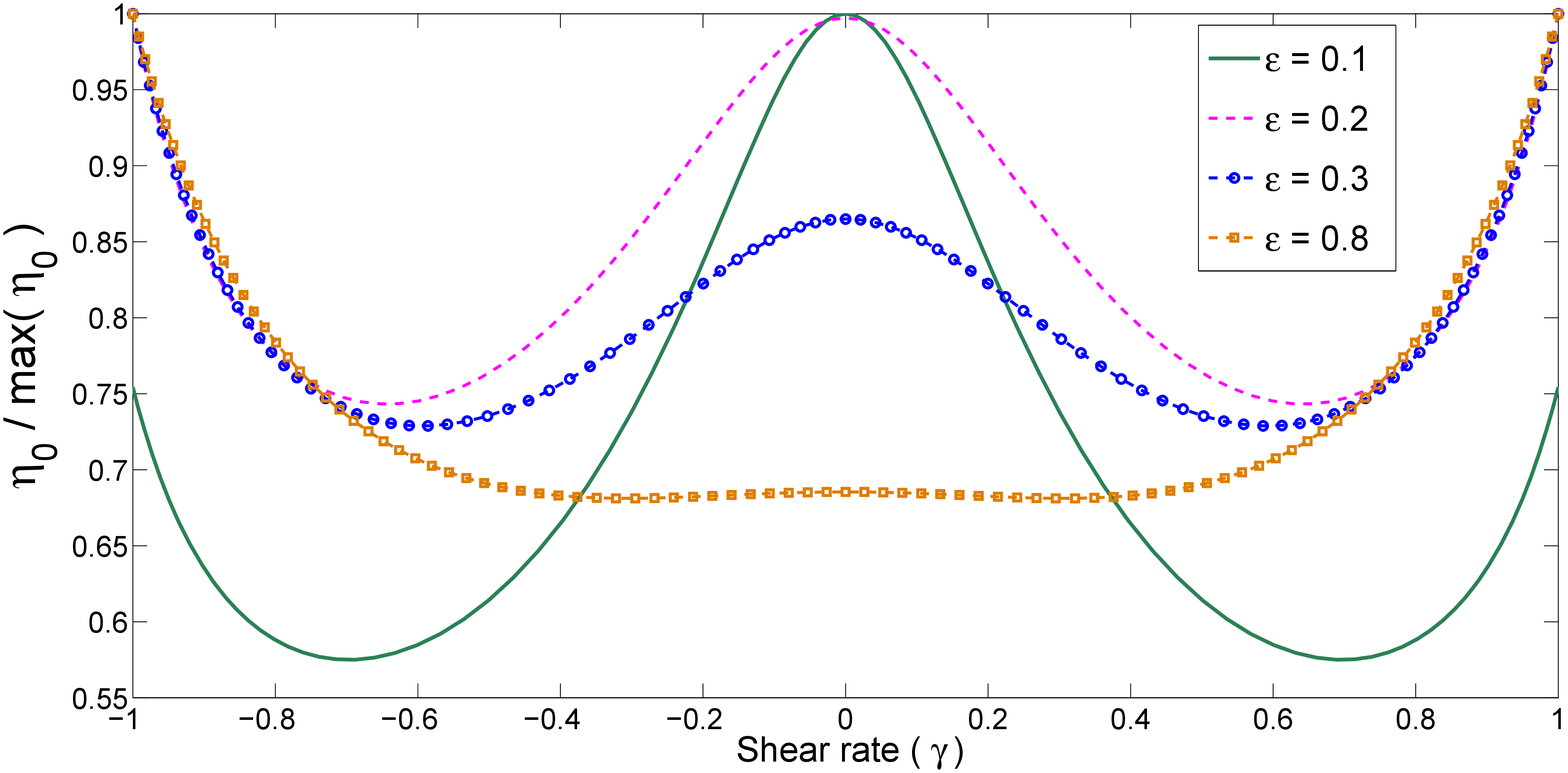}} \\
      \end{tabular}
    \caption{(Color online) In the left figure, equilibrium flow profiles of dusts are plotted for different $\epsilon$. The solid red line curve shows the same for the Newtonian limit($\epsilon,\gamma \rightarrow 0$). In the right figure, non-Newtonian viscosity is plotted against unperturbed velocity shear rate. For $\epsilon = 0.1$, shear thinning property exists until $\gamma = 0.7$ and then shear thickening begins. As $\epsilon$ increases, the property changes from shear thinning to shear thickening and for $\epsilon = 0.8$, shear thinning property almost ceases.}
   \label{Fvis}
  \end{center}
\end{figure}

For the Newtonian viscosity, the solution of the Eq.(\ref{eqlb}) gives a parabolic velocity profile i.e. $v_{0y}(x) = \bar{v}(1-(x/L)^2)$, $L$ is the half width of shear layer. In our analysis, the equilibrium equation(\ref{eqlb}) is solved with the non-Newtonian viscosity model(\ref{model}) for the equilibrium force term $F_0(x)$. Here, \underline{fzero} function of MATLAB is used for solving equation(\ref{eqlb}) to calculate numerically values of $\gamma$ for  each discrete points of space variable $x$ in the range $[-1:1]$. Then the array of values of $\gamma$ is integrated to get equilibrium velocity profile keeping in mind the boundary conditions $v_0 = 0$ for $x = \pm 1$ and $dv_0/dx|_{x=0} = 0$. With the numerical values of $\gamma$, non-Newtonian viscosity could be calculated from the model(\ref{model}). In the figure(\ref{Fvis}), both velocity and corresponding viscosity has been plotted for different values of $\epsilon$. In non-Newtonian regime, flow profiles deviates from the parabolic flow in Newtonian limit. Viscosity is plotted against the shear rate for different values of $\epsilon$ which clearly shows that the fluid property changes from shear thinning to shear thickening with increasing $\epsilon$.
\section{Linear Equations}\label{sec:ana}
We carry out linear stability analysis for the small amplitude wave so that the higher order terms in perturbation can be ignored for the assumption $|v_x|,|v_y| \ll |v_0|$ where $v_x$ and $v_y$ are the components of the small disturbance in dust flow. The total flow is the sum of the equilibrium flow  and a small perturbation in flow:
 \[
 {\bf v}(x,y,t) = [v_0(x) +v_{y}(x,y,t)] \hat e_{y}+   v_{x}(x,y,t)\hat e_{x}.
 \]
Space and time are normalized by the Debye length $\lambda_{D} = \sqrt{ T_i/4 \pi Z n_0 e^2}$ and dust plasma frequency $\omega_{pd}=\sqrt{4 \pi n_0 Z^{2} e^2/m}$ respectively. All densities i.e. electron ion and dust are normalized by $n_{0}$ and the electrostatic potential $\varphi$ by $T_i/e$ ($\phi \equiv e \varphi/T_i$). For small potential fluctuations ($\phi \ll 1$),  the normalized Boltzmann relations(\ref{poiss}) can be expressed  as,
 \[
 n_e = \frac{n_{eo}}{n_{0}}\left( 1 + \phi \frac{T_i}{T_e} \right),  ~~~~~~~~~ n_i = \frac{n_{io}}{n_{0}}\left( 1 - \phi \right).
 \]
 The linearized dimensionless Poisson's equation is written as,
 \bee
 \nabla^2 \phi = n + \alpha \phi,
 \label{lin_pois}
 \ene
where $\alpha = \left(n_{e0} T_i + n_{i0} T_e\right)/(n_0 Z T_e)$.
The linear continuity equation in normalized variables can be written as,
\bee
\left(\fpar{}{t} +v_0 \fpar{}{y} \right)n+ \fpar{v_x}{x} + \fpar{v_y}{y}   = 0.
\label{lin_cont}
\ene
x, y components of the linearized  dimensionless momentum equation of the dust fluid are respectively given by,
\bee
\left(\fpar{}{t} + v_0 \fpar{}{y}\right)v_x  - \fpar{\phi}{x} + c_{d}^2 \fpar{n}{x} =\eta_0 \nabla^2 v_x + \left(\frac{\eta_0}{3} \fpar{}{x} - \frac{2}{3} \eta'_0 v''_0 \right) \left( \nabla \cdot {\bf v} \right)
+ \eta'_0 v'_0 \fpar{}{y} \left( \fpar{v_x}{y}+\fpar{v_y}{x} \right) + 2 \eta'_{0} v''_{0} \fpar{v_x}{x}
\label{vx}
\ene
and
\bee
\left(\fpar{}{t} + v_0 \fpar{}{y}\right)v_y+v_x\fdar{v_0}{x}-\fpar{\phi}{y} + c_{d}^2\fpar{n}{y} = \eta_{0} \nabla^2 v_y + \frac{\eta_0}{3} \fpar{}{y}\left(\nabla \cdot {\bf v}\right)
+ \left\{ 2 \eta'_0 v''_{0} + \eta''_0 v''_0 v'_0 + \eta'_{0} v'_{0} \fpar{}{x} \right\} \left( \fpar{v_x}{y}+\fpar{v_y}{x} \right)
\label{vy}
\ene
where $\eta_0$ is normalized by $\omega_{pd} \lambda_{D}^{2} M n_0$. The $\eta'_0$ and $\eta''_0$ denotes the single and double derivative of $\eta_0$ with respect to $v'_0$ and $v'_0 = dv_0/dx$.
\section{Eigenvalue Analysis}\label{sec:eig}
It is not possible to carry out Fourier analysis along the direction of inhomogeneity. Thus the perturbed variables $v_x$, $v_y$, $\phi$ and $n$ would take the form as $ n(x,y,t) = n(x) e^{i(k y -\omega t)} $. So the linearized equations (\ref{lin_pois}-\ref{vy}) can be expressed as four normalized ordinary differential equation in $y$  by the following equations:
\beea
k v_0(x) n + k v_y - i \fdar{v_x}{x} = \omega n,
\label{pert_con}
\enea
\beea
n + \left( \alpha - \nder{2}{}{x} + k^2 \right) \phi = 0,
\label{pert_pois}
\enea
\beea
-ic_{d}^2 \fdar{n}{x} + i \fdar{\phi}{x}
+ \left[ k v_0 + i \eta_0 \left( \nder{2}{}{x} - k^2\right) - i\eta'_0 v'_0 k^2 + 2 i \eta'_0 v''_0 \fdar{}{x} + i
\left(\frac{\eta_0}{3} \fdar{}{x} - \frac{2}{3}\eta'_0 v''_0\right) \fdar{}{x}\right]v_x\nonumber \\
+ \left[ -\eta'_0 v'_0  \fdar{}{x} - \left(\frac{\eta_0}{3}\fdar{}{x}
- \frac{2}{3}\eta'_0 v''_0\right) \right] k v_y= \omega v_x.
\label{pert_vx}
\enea
\beea
k c_{d}^2 n - k \phi  +\left[ -i\frac{ v'_0}{k} -\left(2 \eta'_0 v''_0 +  \eta''_0 v''_0 v'_0 + \eta'_0v'_0\fdar{}{x}\right)-\frac{\eta_0}{3} \fdar{}{x}\right] kv_x + \nonumber\\
\left[ k v_0 + i \eta_0 \left( \nder{2}{}{x} - k^2 \right)-\frac{i}{3} k^2 \eta_0
+i\left(2 \eta'_0 v''_0 +  \eta''_0 v''_0 v'_0 + \eta'_0v'_0\fdar{}{x}\right)\fdar{}{x} \right] v_y = \omega v_y,
\label{pert_vy}
\enea
\begin{figure}
       \centering
       \includegraphics[width=4in,height=3in]{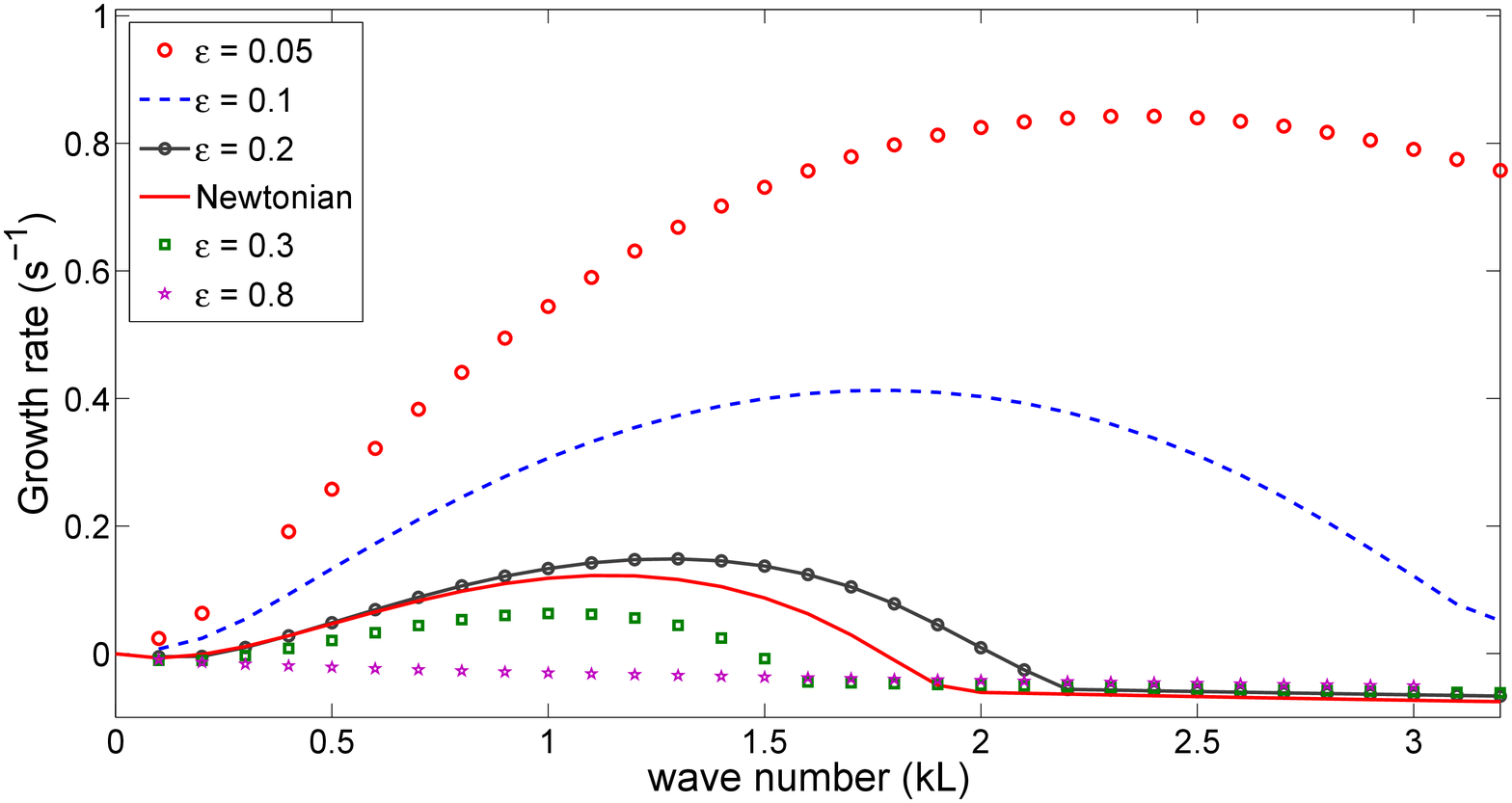}
       \caption{(Color online) Growth rate of instability is plotted against wave number for different values of parameter $\epsilon$ in incompressible limit. The solid red line curve shows that for Newtonian limit. For $\epsilon = 0.2$, the growth rate is close to that of Newtonian limit. }
       \label{incom}
\end{figure}
 We have solved these four coupled linear eigenvalue equations and investigated the growth rate of the KH instability with the variation of different parameters like Mach number($M = |v_0|/c_d$), non-Newtonian parameter($\epsilon$) and wave number($k$).
First we investigate the incompressible limit ($c_d \gg |v_0|$) where the density and the potential fluctuations is negligibly small so that equation(\ref{pert_pois}) becomes trivial and continuity equation(\ref{pert_con}) reduces to $k v_y - i d v_x/d x = 0$. We have carried out the  matrix eigenvalue analysis using the standard eigenvalue subroutine(eig) in MATLAB after proper discretization of the above mentioned equations.  Following central difference scheme has been used for the purpose of discretization
\begin{center}
\beea
\nder{2}{\phi}{x} = \frac{\phi_{i+1}-2\phi_{i}+\phi_{i-1}}{\Delta^2},~~~~~~~~~~~~~~~~~~~\nonumber\\
\nder{}{\phi}{x}  = \frac{\phi_{i+1}-\phi_{i-1}}{2\Delta},~~~~~~~~~~~~~~~~~~~~~~~~~~\nonumber
\enea
\end{center}
where $\Delta$ is the grid spacing. In figure (\ref{incom}), the growth rate is shown plotted against the wave number ($k$) for different values of $\epsilon$. The solid red line curve indicates the Newtonian regime and the other curves  are for values of $\epsilon = 0.05, 0.1, 0.2, 0.3 ~$and$~ 0.8 $. The kinematic viscosity in Newtonian limit is considered as $ 1.538\times10^{-4} m^2/s$. Different values of $\epsilon$ incorporates different functional dependance of viscosity with flow shear rate. In figure(\ref{Fvis}), we have shown how viscosity coefficient changes its property from shear thinning to shear thickening one with increase of plasma temperature $T_0$. As the value of $\epsilon$ is increased to $0.3$,  growth rate diminishes below that of the Newtonian case and the shear thickening property overpowers the effect of shear thinning. Hence, we can summarize that the shear thinning property enhances the instability but on the contrary, the shear thickening property has stabilizing role on the KH instability. For $\epsilon = 0.8$, shear thickening effect stabilizes the medium. Here, variation of viscosity with shear rate plays the stabilizing role on the KH instability.
\begin{figure}
  \begin{center}
    \begin{tabular}{cc}
      \resizebox{90mm}{!}{\includegraphics{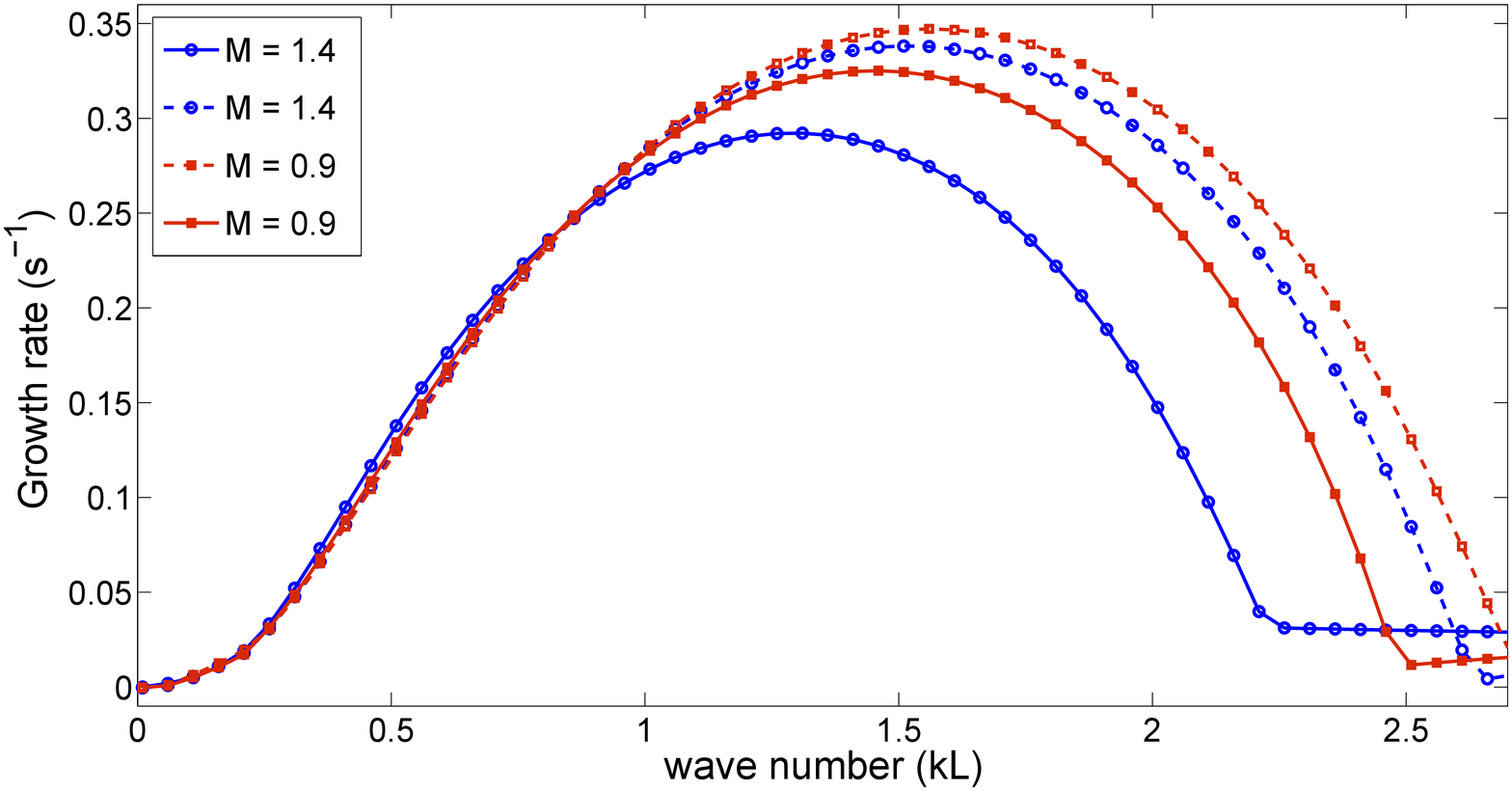}} &
      \resizebox{90mm}{!}{\includegraphics{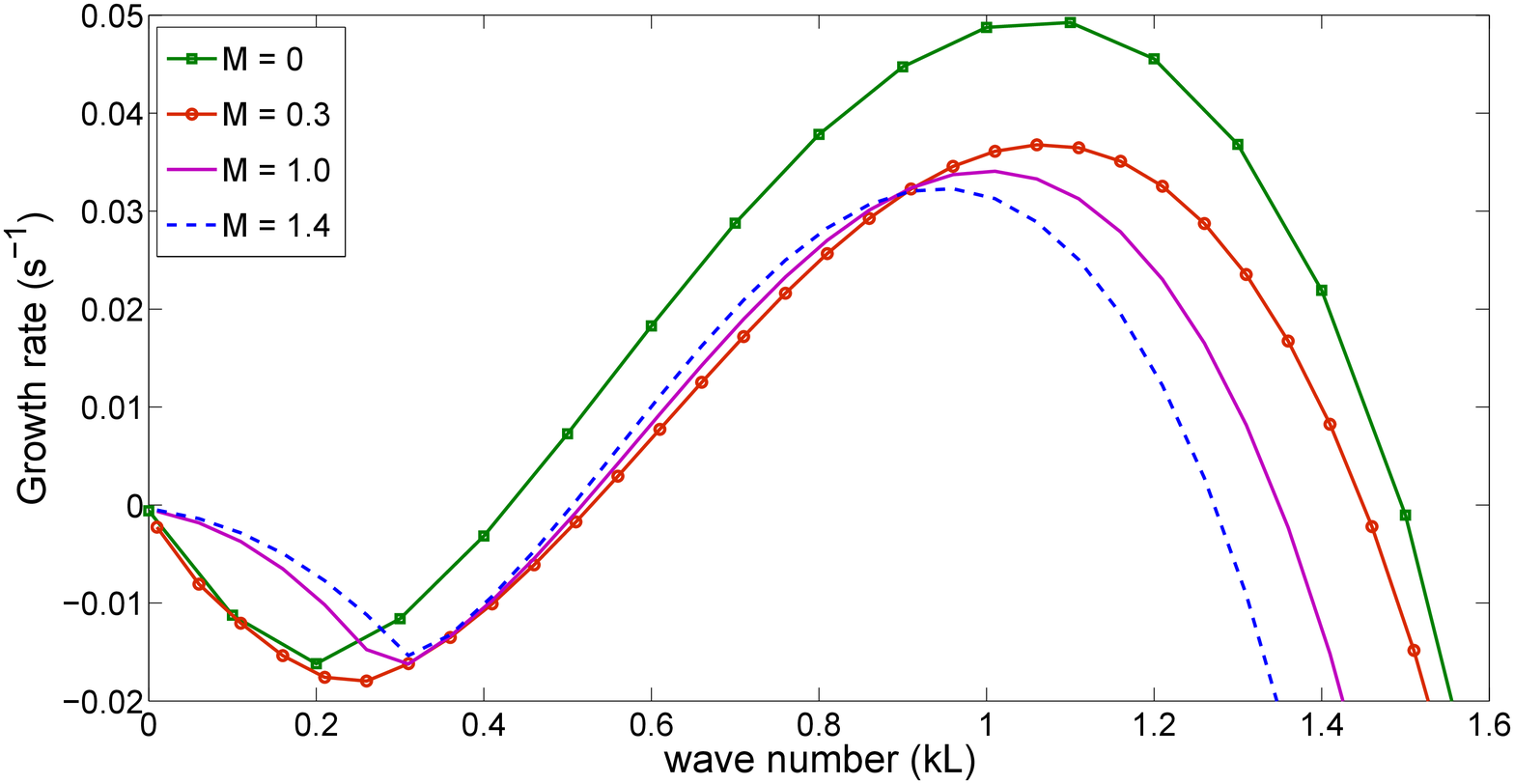}} \\
      \end{tabular}
    \caption{(Color online) In the left figure, two sets of curves are shown for two different Mach number ($M$) including and excluding dispersion term in Poisson's equation for $\epsilon=0.1$. In each set of curves, dotted line represents the curve without dispersion effect and the solid line with dispersion term. In the right one, compressibility is introduced by increasing the Mach number and it indicates that the growth rate diminishes as compressibility strengthens in the medium for $\epsilon = 0.3$. Here, $M = 0$ curves shows incompressible limit for comparison.}
   \label{com}
  \end{center}
\end{figure}
\begin{table*}[h]
\begin{center}{\footnotesize
\begin{tabular}{| c | c | c | l |}
\hline
~~~ kL ~~~ & ~~~~~~~~$\epsilon$ ~~~~~~~&~~~~~~$M$ ~~~~~~~  &~~~ Growth rate($s^{-1}$)  \\
\hline
 $  $  &    $  $ &   $ 0.5 $    &  $~~~~~~~~~ 0.3258 $ \\
 $  $  & $  $ &  $ 0.7 $    &    $~~~~~~~~~  0.3215 $ \\
 $  $  &    $  $ &   $ 0.9 $   &  $~~~~~~~~~  0.3146 $ \\
 $1.2  $  &    $0.1  $ &   $ 1.2 $    &    $~~~~~~~~~  0.3032 $ \\
 $  $  &    $  $ &   $ 1.4 $    &    $~~~~~~~~~  0.2902 $ \\
 $  $  &    $  $ &   $ 2.0 $    &    $~~~~~~~~~  0.2542 $ \\
 $  $  &    $  $ &   $ 2.5 $    &    $~~~~~~~~~  0.2179 $ \\
\hline
$  $   &    $  $ &   $ 0.5 $    &  $~~~~~~~~~ 0.1208 $ \\
 $  $  & $  $ &  $ 0.7 $    &    $~~~~~~~~~  0.1178 $ \\
 $  $  &    $  $ &   $ 0.9 $   &  $~~~~~~~~~  0.1137 $ \\
 $1.2$ & $0.2  $ &   $ 1.2 $    &    $~~~~~~~~~  0.1056 $ \\
 $  $  &    $  $ &   $ 1.4 $    &    $~~~~~~~~~  0.0989 $ \\
 $  $  &    $  $ &   $ 2.0 $    &    $~~~~~~~~~  0.0623 $ \\
 $  $  &    $  $ &   $ 2.5 $    &    $~~~~~~~~~  0.0368 $ \\
\hline
\end{tabular}}
\end{center}
\caption{ Comparison of growth rates for different parameters $M$, $k$ and $\epsilon$.}
\label{table:cmpre}
\end{table*}

\begin{figure}
  \begin{center}
    \begin{tabular}{cc}
      \resizebox{80mm}{!}{\includegraphics{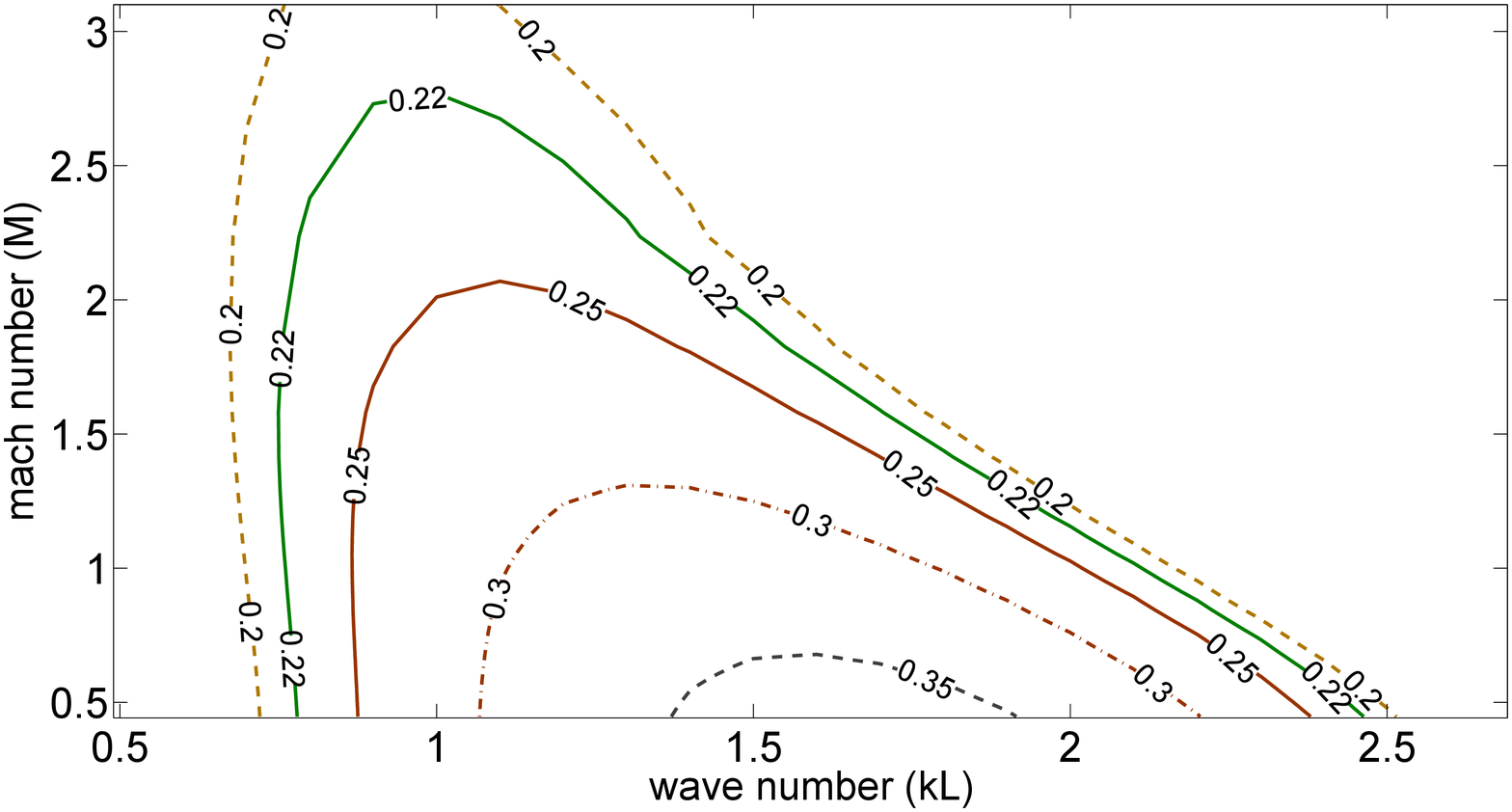}} &
      \resizebox{95mm}{!}{\includegraphics{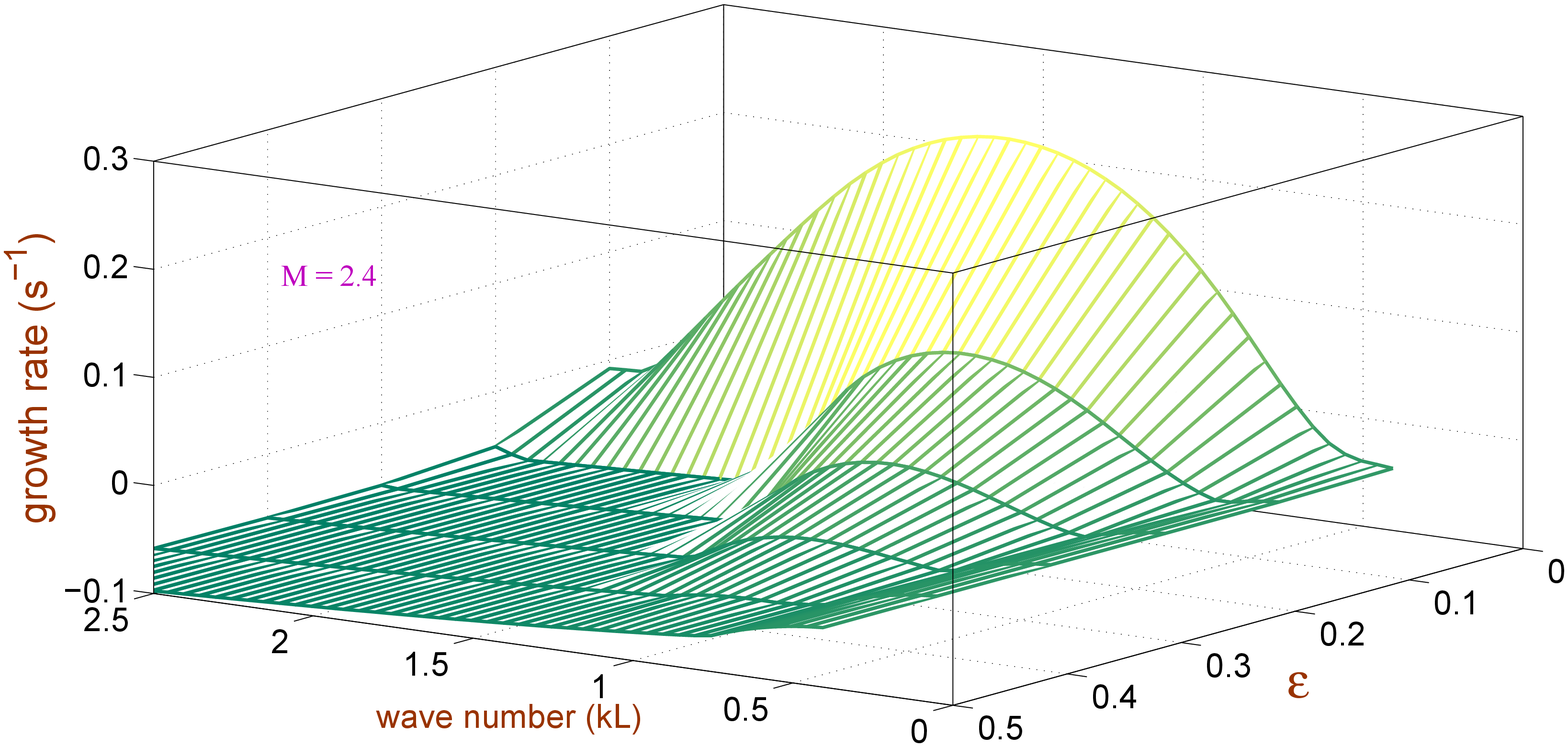}} \\
      \end{tabular}
    \caption{(Color online) The left figure shows contour plot of growth rate in the plane of  mach number ($M$) and wave number for $\epsilon = 0.1$. In the right figure surface plot of growth rate is drawn on the parametric space of $\epsilon$ and $k$ for mach no. $M=2.4$. }
   \label{3d}
  \end{center}
\end{figure}

Now, we include the effects of compressibility in our system to study the role of density fluctuation on the instability.  Figure(\ref{com}) shows that the growth rate decreases as we increase the Mach number i.e., compressibility weakens the instability. Inclusion of compressibility effect enables dissipation of  some energy to drive longitudinal waves. For small mach no, compressibility effect is too weak to stabilize but here, shear thickening property could play the role. In plasma, quasi-neutrality is a widely accepted approximation for wavelengths larger than Debye length ($\lambda_{D}$) where the dispersion term of poisson's equation has negligible contribution. In figure(\ref{com}), for two different values of $M = 1.4, 0.9$ growth rate is plotted with and without considering the dispersion term in Poisson's equation. The dispersion is much prominent for higher compressibility ($M = 1.4$). So, in the regime of higher mach number quasi-neutrality is not a correct approximation. In figure (\ref{3d}), contour plot of growth rate is drawn in 2D plane of wave number and Mach number and it is seen that as growth rate decreases from $0.3$ to $0.21$, the unstable region spans. A surface plot of the growth rate vs wave number and $\epsilon $ is shown for $M = 2.4$. The flat area addresses the stability region on the $(\epsilon-k)$ plane and the hill area indicates unstable portion. For higher temperature (large values of $\epsilon$), the unstable region shrinks and flat area widens which depicts the stabilizing effect of shear thickening property.
%
\section{Summary}\label{sec:sum} %
To summarize, in this article we have investigated stability of inhomogeneous bounded flow in non-Newtonian dusty plasma. With an experimentally justified mathematical model of variation of viscosity with equilibrium velocity shear rate, the linear stability analysis is carried out numerically using standard matrix eigenvalue technique. In figure(\ref{Fvis}), viscosity coefficient and corresponding equilibrium velocity flow profiles are shown which are the solutions of the equilibrium momentum balance equation(\ref{eqlb}). Depending on the variable parameter $\epsilon$ (depends on the ratio of equilibrium plasma temperature to dust crystal melting temperature), non-Newtonian property changes from shear thinning to shear thickening regime. Here, we have shown that shear thinning property is more favorable for Kelvin-Helmholtz instability so that small values of $\epsilon$ increases the growth rate which is more strong compared with the Newtonian limit. On the shear thickening regime (large values of $\epsilon$), growth rate diminishes and hence it acts against the instability and stabilizes the medium. For the KH instability, incompressible limit shows the maximum growth of instability and as we include finite density fluctuation (compressibility), a part of energy available for the instability is exhausted for the longitudinal fluctuation in the system and thus instability weakens by some percentage.
Viscosity has dissipative effect in fluid but also it has the nature of diffusing momentum. In bounded flow strong velocity shear exists in boundary layers which is diffused outwards by viscosity which leads to instability\cite{dzre}. In our case, equilibrium shear rate increases with decrease of parameter $\epsilon$ near the boundary and accordingly for lower values of $\epsilon$ growth rate is much greater than that of Newtonian limit.

Before concluding, we make comments on the effect of dust charge fluctuation and strong coupling between dust particles on the KH instability. Since charge is a dynamical variable in dusty plasmas, a complete analysis should involve dust charge fluctuations. However, such effects are known to produce the usual damping on the dynamics of dust particles\cite{srb,jana}. Dust charge fluctuations result from  perturbation in electron and ion densities. So, the effect of dust charge fluctuation does not manifest itself in the incompressible limit. But, for compressible dusty plasma, dust charge fluctuation gives rise to some damping effect on the KH instability.
In our present analysis, the strong coupling effect between dust particles is not included. However, in our earlier publication \cite{baner} it has been shown that the strong coupling introduces elastic property in dusty plasma which usually enhances the KH instability. So, study of the KH instability in strongly coupled non-Newtonian dusty plasma would be interesting and is therefore left for future work.

As amplitude of small velocity fluctuation increases due to instability, system goes to nonlinear state and it would start to show vortex formation due to the convective nonlinearity of momentum equation. In non-Newtonian system, viscous stress tensor drives another type of nonlinear effect through the dependence of viscosity on shear flow rate and it could lead to many interesting phenomena like recurrence \cite{dbnl}. In nonlinear regime, it would be interesting to study these nonlinear effects on KH instability.

%
%


   \end{document}